# On the very accurate evaluation of the Voigt/complex error function with small imaginary argument

Yihong Wang

School of Energy and Environment, Southeast University, Nanjing 210096, China

*wyh@seu.edu.cn*

**Abstract** A rapidly convergent series, based on Taylor expansion of the imaginary part of the complex error function, is presented for highly accurate approximation of the Voigt/complex error function with small imaginary argument $y \leq 0.1$. Error analysis and run-time tests in double-precision computing platform reveals that in the real and imaginary parts the proposed algorithm provides average accuracy exceeding $10^{-15}$ and $10^{-16}$, respectively, and the calculation speed is as fast as that of reported in recent publications. An optimized MATLAB code providing rapid computation with high accuracy is presented.

## 1. Introduction

The complex error function also known as the Faddeeva function is given by[1, 2]

$$w(z) = e^{-z^2} \text{erfc}(-\mathrm{i}\, z)$$
$$= e^{-z^2}(1 + \frac{2\mathrm{i}}{\sqrt{\pi}} \int_0^z e^{-t^2} dt), \tag{1}$$

where $z = x + iy$ is the complex argument and $y > 0$. Using the Fourier transforms, the real and imaginary parts of the complex error function (1) can be represented as[3]

$$K(x, y) = \frac{1}{\sqrt{\pi}} \int_0^{+\infty} \exp(-\frac{1}{4}t^2) \exp(-yt) \cos(xt) dt, \tag{2}$$

and

$$L(x, y) = \frac{1}{\sqrt{\pi}} \int_0^{+\infty} \exp(-\frac{1}{4}t^2) \exp(-yt) \sin(xt) dt, \tag{3}$$

respectively. The real part of the complex error function $K(x, y)$ is known as the Voigt function that are widely used in many fields of Physics, Chemistry and Astronomy[4-6].

Because there is no closed-form solution for the integrals above, many modern "state-of-the art" algorithms for evaluating the Voigt/complex error function utilizing sophisticated numerical techniques have been discussed in numerous papers. Several highly accurate algorithms[7-9] or arbitrary precision algorithms[10, 11] have been proposed to give the benchmark value. However, due to the high cost of computation, these algorithms are not suitable for large-scale computational applications such as high resolution line-by-line radiative transfer modeling[12]. Accurate yet efficient computation of the Voigt and complex error function is a challenge since decades in astrophysics and other areas of physics.

Rational approximations are particularly attractive and are used in many algorithms, because they can be implemented efficiently and allow for high accuracy[13]. The rational approximations algorithms "cpf12"[14] and "w4"[15] proposed by Humlicek appear to belong to the most popular complex error function algorithms and several modified algorithms[16-18] on the basis of Humlicek's algorithms have been further developed to improve the calculation accuracy or expand the scope of application. Humlicek's algorithms and its modified algorithms provide accuracy chosen between $10^{-2}$ and $10^{-6}$ for almost the entire complex plane except for very small imaginary values of the argument. Recently, several high-accuracy algorithms such as "fexp"[19, 20], "voigtf"[21] and "fadsamp"[22] algorithms proposed by Abrarov and his collaborators were developed to achieve highly accurate and simultaneously rapid computation of Voigt/complex error function. The accuracy of the "fexp" algorithm based on Fourier expansion of the exponential multiplier for the real and imaginary parts of the complex error function are $10^{-9}$ and $10^{-8}$, respectively, in Humlicek regions 3 and 4. The accuracy of the "voigtf" algorithm with 16 summation terms of rational fraction for the Voigt function is better than $10^{-8}$ in the domain of $y > 10^{-6}$. However, the accuracy of both "fexp" algorithm and "voigtf" algorithm deteriorate significantly with decreasing $y$. Compared with the former two algorithms, the "fadsamp" algorithm based on incomplete cosine expansion of the sinc function sustains high accuracy ($\sim 10^{-13}$) in computation at smaller values of the parameter $y$ that is commonly considered difficult for computation of the Voigt/complex error function. However, the "fadsamp" algorithm needs at least 24 terms to achieve the accuracy of $10^{-13}$, which is not conducive to fast calculation.

According to the above analysis, almost all algorithms will face the challenge of accurate calculation of Voigt/complex error function with small $y$. Recently, Abrarov and Quine[23] proposed an efficient algorithm (the accuracy is better than $10^{-12}$ while $y < 10^{-6}$) based on Maclaurin expansion of the exponential function to overcome the notoriously difficult. Although these algorithms are sufficient for the most practical tasks, the more accurate and simultaneously rapid calculation of the Voigt/complex error function may also be required.

In this work we propose a new algorithm for the Voigt/complex error function with small imaginary argument ($y \leq 0.1$) based on Taylor series expansion method. By adding a finite term, this algorithm can be easily extended to arbitrary precision. We applied MATLAB R2019a supporting array programming features to implement the numerical verification of this algorithm, and a typical desktop computer Intel(R) Quad CPU with RAM 8.00 GB was utilized.

## 2. Theory and methods

2.1. Evaluation for the imaginary part $L(x, y)$

The $n$-order partial derivatives of $L(x, y)$ with respect to $y$ can be written as follow

$$\frac{\partial^n L(x,y)}{\partial y^n} = \frac{1}{\sqrt{\pi}} \int_0^{+\infty} \exp(-\frac{1}{4}t^2) \frac{d^n \exp(-yt)}{dy^n} \sin(xt)dt$$
$$= \frac{(-1)^n}{\sqrt{\pi}} \int_0^{+\infty} t^n \exp(-\frac{1}{4}t^2 - yt)\sin(xt)dt. \tag{4}$$

By substituting $y = 0$ into the integral above, the $n$-order partial derivatives along the $x$ axis yields

$$\frac{\partial^n L(x,0)}{\partial y^n} = \frac{(-1)^n}{\sqrt{\pi}} \int_0^{+\infty} t^n \exp(-\frac{1}{4}t^2)\sin(xt)dt. \tag{5}$$

While $n$ is an odd number, the integral (5) has an analytic solution as follow[24]

$$\frac{\partial^n L(x,0)}{\partial y^n} = (-1)^{(n+1)/2} \exp(-x^2) H_n(x) \qquad \text{for } n \text{ odd}, \tag{6}$$

where $H_n(x)$ represents the $n$-order Hermite polynomials of real argument $x$:

$$H_n(x) = h_{n,0}x + h_{n,1}x^3 + \ldots + h_{n,(n-1)/2}x^n = \sum_{k=0}^{(n-1)/2} h_{n,k} x^{2k+1}, \tag{7}$$

where the coefficients are

$$h_{n,k} = (-1)^{\frac{n-1}{2}-k} \frac{n!}{((n-1)/2-k)!(2k+1)!} 2^{2k+1}. \tag{8}$$

There is no closed-form solution for the integral (5) while n is an even number, however, we have proved that the even-order partial derivatives can be expressed in terms of the Dawson's integral $D(x)$ of real argument x as follow (Appendix A)

$$\frac{\partial^n L(x,0)}{\partial y^n} = \frac{1}{\sqrt{\pi}} (P_{n/2}(x)D(x) + Q_{n/2}(x)) \qquad \text{for } n \text{ even,} \tag{9}$$

where the Dawson's integral $D(x)$ is defined as

$$D(x) = \frac{1}{2}\int_0^{+\infty} \exp(-\frac{1}{4}t^2)\sin(xt)dt, \tag{10}$$

and $P_{n/2}(x)$, $Q_{n/2}(x)$ are polynomials of real argument $x$:

$$P_{n/2}(x) = p_{n/2,0} + p_{n/2,1}x^2 + \ldots + p_{n/2,n/2}x^n = \sum_{k=0}^{n/2} p_{n/2,k}x^{2k}, \tag{11}$$

and

$$Q_{n/2}(x) = q_{n/2,0}x + q_{n/2,1}x^3 + \ldots + q_{n/2,n/2-1}x^{n-1} = \sum_{k=0}^{n/2-1} q_{n/2,k}x^{2k+1}, \tag{12}$$

where the coefficients $p_{n/2,k}$ and $q_{n/2,k}$ can be derived from the following recurrence relations

$$P_m(x) = \begin{cases} 2 & m = 0 \\ 4 - 8x^2 & m = 1 \\ (8m - 6 - 4x^2)P_{m-1}(x) - 8(m-1)(2m-3)P_{m-2}(x) & m \geq 2, \end{cases} \tag{13}$$

and

$$Q_m(x) = \begin{cases} 0 & m = 0 \\ 4x & m = 1 \\ (8m - 6 - 4x^2)Q_{m-1}(x) - 8(m-1)(2m-3)Q_{m-2}(x) & m \geq 2, \end{cases} \tag{14}$$

respectively. It should be noted that the calculation of the Dawson's integral $D(x)$ is not difficult and several efficient approximations that can provide rapid and highly accurate computation are reported in literature[25-27]. In this work, a finite continued fraction is used for efficient calculation of Dawson's integral[27]:

$$D(x) \approx F(x, N_D) = \frac{x}{1+2x^2 -} \frac{4x^2}{3+2x^2 -} \frac{8x^2}{5+2x^2 -} \frac{12x^2}{7+2x^2 -} \cdots \frac{4N_D x^2}{2N_D + 1 + 2x^2}. \tag{15}$$

Consequently, the imaginary part $L(x, y)$ of complex error function can be represented as the following Taylor expansion series near $y = 0$:

$$\begin{aligned} L(x,y) &= \sum_{n=0}^{\infty} \frac{1}{n!}\frac{\partial^n L(x,0)}{\partial y^n} y^n \\ &\approx \sum_{n=0}^{\infty} [\frac{1}{(2n)!\sqrt{\pi}}(P_n(x)F(x,M) + Q_n(x))y^{2n} + \frac{(-1)^{n+1}}{(2n+1)!}e^{-x^2}H_{2n+1}(x)y^{2n+1}]. \end{aligned} \tag{16}$$

By calculating the first $N+1$ terms of Eq. (16) and extracting the common factors $D(x)$ and $\exp(-x^2)$, which is the "slowest" array for computation, outside the formula, an efficient approximation for the imaginary part $L(x, y)$ is yielded as follow

$$L(x,y) \approx \frac{1}{\sqrt{\pi}} F(x,N_D) \sum_{n=0}^{N} \alpha_n x^{2n} + xe^{-x^2} \sum_{n=0}^{N} \beta_n x^{2n} + \frac{1}{\sqrt{\pi}} x \sum_{n=0}^{N-1} \gamma_n x^{2n}, \tag{17}$$

where the coefficients are

$$\begin{aligned} \alpha_n &= \sum_{m=n}^{N} \frac{1}{(2m)!} p_{m,n} y^{2m}, \\ \beta_n &= \sum_{m=n}^{N} \frac{(-1)^{m+1}}{(2m+1)!} h_{2m+1,n} y^{2m+1}, \\ \gamma_n &= \sum_{m=n+1}^{N} \frac{1}{(2m)!} q_{m,n} y^{2m}. \end{aligned} \tag{18}$$

2.2. Evaluation for the real part $K(x, y)$

The 1-order partial derivatives of $L(x, y)$ with respect to $y$ can be written as follow

$$\frac{\partial L(x, y)}{\partial y} = \frac{-1}{\sqrt{\pi}} \int_0^{+\infty} t \exp(-\frac{1}{4}t^2 - yt) \sin(xt) dt.$$

$$= \frac{2y}{\sqrt{\pi}} \int_0^{\infty} \exp(-\frac{t^2}{4}) \exp(-yt) \sin(xt) dt - \frac{2x}{\sqrt{\pi}} \int_0^{\infty} \exp(-\frac{t^2}{4}) \exp(-yt) \cos(xt) dt \quad (19)$$

$$= 2yL(x, y) - 2xK(x, y).$$

Thus the real part $K(x, y)$ of complex error function can be represented as

$$K(x, y) = \frac{y}{x} L(x, y) - \frac{1}{2x} \frac{\partial L(x, y)}{\partial y}. \quad (20)$$

Consequently, an efficient approximation for $K(x, y)$ is obtained by substituting the Eq. (17) into the Eq. (20):

$$K(x, y) \approx \begin{cases} \frac{1}{\sqrt{\pi}} \frac{1}{x} F(x, N_D) \sum_{n=0}^{N} \alpha'_n x^{2n} + e^{-x^2} \sum_{n=0}^{N} \beta'_n x^{2n} + \frac{1}{\sqrt{\pi}} \sum_{n=0}^{N-1} \gamma'_n x^{2n} & \text{for } x > 0 \\ \frac{1}{\sqrt{\pi}} \alpha'_0 + e^{-x^2} \beta'_0 + \frac{1}{\sqrt{\pi}} \gamma'_0 & \text{for } x = 0, \end{cases} \quad (21)$$

where the coefficients $\alpha'_n$, $\beta'_n$ and $\gamma'_n$ are

$$\alpha'_n = y\alpha_n - \frac{1}{2} \sum_{m=n}^{N} \frac{\text{sign}(m)}{(2m - \text{sign}(m))!} p_{m,n} y^{2m-1},$$

$$\beta'_n = y\beta_n - \frac{1}{2} \sum_{m=n}^{N} \frac{(-1)^{m+1}}{(2m)!} h_{2m+1,n} y^{2m}, \quad (22)$$

$$\gamma'_n = y\gamma_n - \frac{1}{2} \sum_{m=n+1}^{N} \frac{1}{(2m-1)!} q_{m,n} y^{2m-1}.$$

## 3. Algorithmic implementation

3.1 Evaluation scheme

Many previous studies have claimed that when the argument $z$ is large enough by absolute value, the truncation of the Laplace continued fraction[7, 8]

$$w(z) \approx C(z, N_C) = \frac{i/\sqrt{\pi}}{z} \frac{1/2}{-z} \frac{1}{-z-} \frac{3/2}{z} \frac{}{-} \cdots \frac{N_C/2}{-z}, \quad (23)$$

can be effectively used for high-accuracy and rapid computation of the complex error function $w(z)$. However, we note that when $y$ is quite small, the convergence speed of Laplace continued fraction is significantly reduced, even if $|z|$ is large enough, which require a lot of calculation, leading to a serious decline in computational efficiency. For example, the approximation accuracy of Eq. (23) with $y = 1 \times 10^{-20}$ at different $N_C$ is as shown in Fig. 1a, in which the relative errors of Eq. (23) is defined as follow

$$\Delta_C(|z|, N_C) = \max\{\frac{|\text{Re}[C(z, N_C)] - \text{Re}[w_{\text{ref.}}(z)]|}{\text{Re}[w_{\text{ref.}}(z)]}, \frac{|\text{Im}[C(z, N_C)] - \text{Im}[w_{\text{ref.}}(z)]|}{\text{Im}[w_{\text{ref.}}(z)]}\}, \quad (24)$$

where the highly accurate reference values of $w_{\text{ref.}}(z)$ can be obtained according to Eq. (1) by using the MATLAB that supports error function of complex argument.

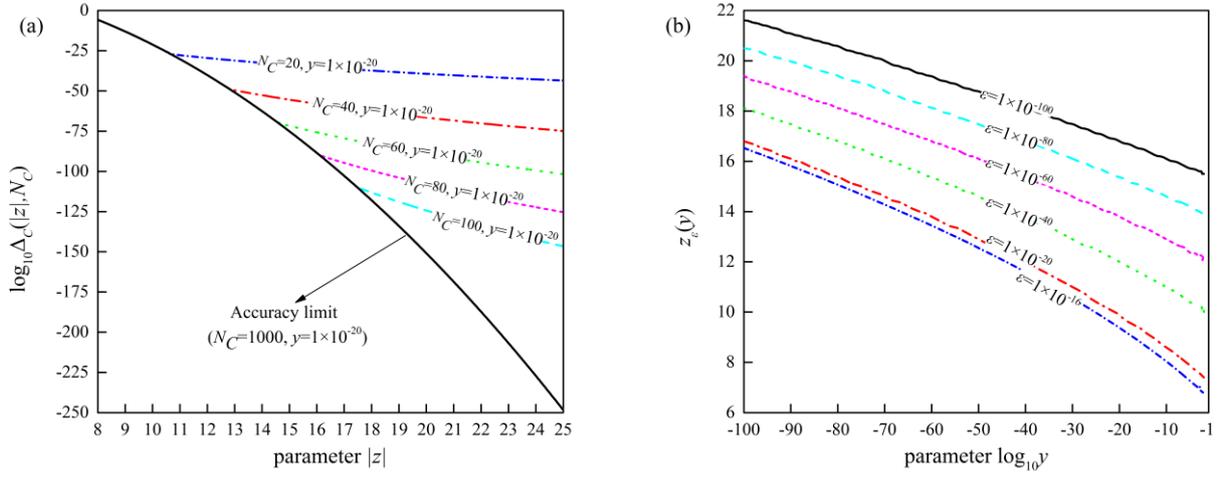

Fig. 1. (a) The approximation accuracy of Eq.(23) with $y = 1\times10^{-20}$ at different $N_C$ and (b) the efficient computing boundary $z_\varepsilon(y)$ with different selected accuracy.

As we can see from Fig. 1a, the relative errors at $|z| = 8$ is still greater than $1\times10^{-6}$ even if $N_C$ is increased to 1000. Fortunately, the approximation accuracy of Eq. (23) with $y = 1\times10^{-20}$ increases significantly with increasing $|z|$, i.e. only the first 100 terms need to be taken, and the truncation of the Laplace continued fraction can reach the approximation accuracy of $10^{-100}$ when $|z| > 16.8$. Therefore, it is important to determine the efficient computing boundary of the Laplace continued fraction approximations, especially when $y$ is small. The numerical calculation results illustrate that if it is desired to achieve the selected precision $\varepsilon$ efficiently, the absolute value of argument $z$ needs to be greater than a certain boundary $z_\varepsilon(y)$ which is determined by $y$ and $\varepsilon$ together and can be written as follow

$$z_\varepsilon(y) = \min\{|z| \,\|\, \Delta_C(|z|, N_C) \leq \varepsilon\}. \tag{25}$$

Fig. 1b shows the $z_\varepsilon(y)$ curves with selected accuracy of $\varepsilon = 10^{-16}$, $10^{-20}$, $10^{-40}$, $10^{-60}$, $10^{-80}$ and $10^{-100}$, respectively, and the efficient computing boundary in the domain $y \leq 0.1$ can be approximately calculated with the following formulas in Table 1. Thus, in order to improve the calculation efficiency, we use the following scheme to calculate the Voigt/complex error function with small imaginary argument:

$$w(x, y \leq 0.1) \approx \begin{cases} \text{Eq.(17) \& Eq.(21)} & \text{for } |x+iy| < z_\varepsilon(y) \\ \text{Eq.(23)} & \text{for } |x+iy| \geq z_\varepsilon(y). \end{cases} \tag{26}$$

In addition, Fig. 1b suggests that when only the calculation of the complex error function in the domain of $10^{-100} \leq y \leq 0.1$ is considered, and the selected accuracy $\varepsilon$ is not less than $10^{-100}$, the following calculation scheme can be adopted:

$$w(x, 10^{-100} \leq y \leq 0.1) \approx \begin{cases} \text{Eq.(17) \& Eq.(21)} & \text{for } |x+iy| < 22 \\ \text{Eq.(23)} & \text{for } |x+iy| \geq 22. \end{cases} \tag{27}$$

Table 1. Approximation formulas for $z_\varepsilon(y)$ ($y \leq 0.1$)

| $\varepsilon$ | $z_\varepsilon(y)$ |
|---|---|
| $10^{-16}$ | $6.4908-6.9856\times10^{-2}\times\ln y-1.8237\times10^{-4}\times\ln^2 y-3.0026\times10^{-7}\times\ln^3 y$ |
| $10^{-20}$ | $7.1461-6.5589\times10^{-2}\times\ln y-1.6308\times10^{-4}\times\ln^2 y-2.6500\times10^{-7}\times\ln^3 y$ |
| $10^{-40}$ | $9.8625-5.0156\times10^{-2}\times\ln y-9.3640\times10^{-5}\times\ln^2 y-1.3861\times10^{-7}\times\ln^3 y$ |
| $10^{-60}$ | $11.9611-4.2288\times10^{-2}\times\ln y-6.5582\times10^{-5}\times\ln^2 y-9.4912\times10^{-8}\times\ln^3 y$ |
| $10^{-80}$ | $13.7687-3.6042\times10^{-2}\times\ln y-3.6111\times10^{-5}\times\ln^2 y-3.1788\times10^{-8}\times\ln^3 y$ |
| $10^{-100}$ | $15.3784-3.1655\times10^{-2}\times\ln y-1.9984\times10^{-5}\times\ln^2 y-2.0282\times10^{-9}\times\ln^3 y$ |

## 3.2 Parameter optimization

In general, the accuracy of evaluation scheme (26) or (27) will be improved with the increase of parameters $N$, $N_D$ and $N_C$. Numerical calculations (see Fig. 2a) show that the significant number of the Dawson's integral calculated by Eq. (15) increases linearly with increasing $N_D$, which indicates that the Dawson's integral can efficiently perform arbitrary precision calculations. However, we note that the evaluation scheme (26) or (27) has a limit accuracy related to $y$, even if the error caused by the Dawson's integral is ignored (see Fig. 2b). Taking $y = 10^{-4}$ as an example, the significant number of the Voigt/complex error function increases linearly with increasing $N$ when $N$ is less than 12. In contrast, when $N$ is greater than 12, the calculation efficiency is significantly reduced. Numerical calculations suggest that greater $N$, $N_D$ and $N_C$ are unnecessary in some cases, we may select the optimal parameters to minimize the number of terms in the evaluation scheme (26) or (27) in order to gain computational acceleration. In this work, the optimal parameters under different accuracy levels $\Delta=1\times10^{-100}$ and $\Delta=1\times10^{-16}$ of evaluation scheme (27) are shown in Table 2 and Table 3, respectively.

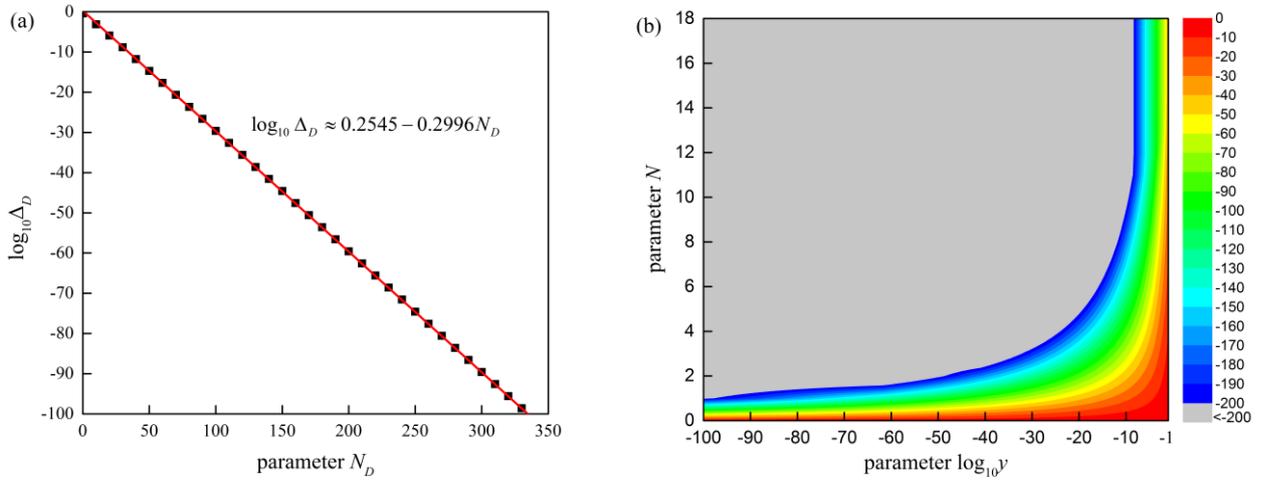

Fig. 2. (a) The maximum relative error $\Delta_D$ of Dawson's integral calculated by Eq.(15) in the domain $0 \leq x < 22$ under different $N_D$ and (b) the maximum relative error $\Delta_w$ of Voigt/complex error function calculated by Eq.(27) in the domain $|x + iy| < 22$ and $10^{-100} \leq y \leq 0.1$ under different $N$ and $y$ with the error caused by the Dawson's integral is ignored ($N_D = 1000$).

Table 2. The optimal parameters $N$, $N_D$, and $N_C$ with accuracy levels $\Delta = 1\times10^{-100}$.

| $y$ | $N$ | $N_D$ | $N_C$ | $\Delta$ |
|---|---|---|---|---|
| $1\times10^{-100} \leq y < 6.3096\times10^{-49}$ | 1 | 344 | 65 | $< 1\times10^{-100}$ |
| $6.3096\times10^{-49} \leq y < 1.5849\times10^{-24}$ | 2 | 344 | 65 | $< 1\times10^{-100}$ |
| $1.5849\times10^{-24} \leq y < 1.5849\times10^{-16}$ | 3 | 344 | 65 | $< 1\times10^{-100}$ |
| $1.5849\times10^{-16} \leq y < 1.5849\times10^{-12}$ | 4 | 344 | 65 | $< 1\times10^{-100}$ |
| $1.5849\times10^{-12} \leq y < 3.9811\times10^{-10}$ | 5 | 344 | 65 | $< 1\times10^{-100}$ |
| $3.9811\times10^{-10} \leq y < 1.5849\times10^{-8}$ | 6 | 344 | 65 | $< 1\times10^{-100}$ |
| $1.5849\times10^{-8} \leq y < 1.5849\times10^{-7}$ | 7 | 344 | 65 | $< 1\times10^{-100}$ |
| $1.5849\times10^{-7} \leq y < 1.5849\times10^{-6}$ | 8 | 344 | 65 | $< 1\times10^{-100}$ |
| $1.5849\times10^{-6} \leq y < 6.3096\times10^{-6}$ | 9 | 344 | 65 | $< 1\times10^{-100}$ |
| $6.3096\times10^{-6} \leq y < 2.5119\times10^{-5}$ | 10 | 344 | 65 | $< 1\times10^{-100}$ |
| $2.5119\times10^{-5} \leq y < 6.3096\times10^{-5}$ | 11 | 344 | 65 | $< 1\times10^{-100}$ |
| $6.3096\times10^{-5} \leq y < 1.5849\times10^{-4}$ | 12 | 344 | 65 | $< 1\times10^{-100}$ |
| $1.5849\times10^{-4} \leq y < 3.9811\times10^{-3}$ | 13 | 254 | 43 | *$7.7625\times10^{-74}$ |
| $3.9811\times10^{-3} \leq y < 0.025119$ | 14 | 197 | 30 | *$4.8978\times10^{-57}$ |
| $0.025119 \leq y < 0.063096$ | 15 | 169 | 25 | *$1.2023\times10^{-48}$ |
| $0.063096 \leq y \leq 0.1$ | 16 | 154 | 22 | *$1.9055\times10^{-44}$ |

* represents the limit accuracy that can be achieved.

Table 3. The optimal parameters $N$, $N_D$, and $N_C$ with accuracy levels $\Delta = 1\times10^{-16}$.

| $y$ | $N$ | $N_D$ | $N_C$ | $\Delta$ |
|---|---|---|---|---|
| $1\times10^{-100} \leq y < 1\times10^{-7}$ | 1 | 61 | 6 | $< 1\times10^{-16}$ |
| $1\times10^{-7} \leq y < 2.5119\times10^{-4}$ | 2 | 61 | 6 | $< 1\times10^{-16}$ |
| $2.5119\times10^{-4} \leq y < 3.9811\times10^{-3}$ | 3 | 61 | 6 | $< 1\times10^{-16}$ |
| $3.9811\times10^{-3} \leq y < 0.015849$ | 4 | 61 | 6 | $< 1\times10^{-16}$ |
| $0.015849 \leq y < 0.039811$ | 5 | 61 | 6 | $< 1\times10^{-16}$ |
| $0.039811 \leq y < 0.063096$ | 6 | 61 | 6 | $< 1\times10^{-16}$ |
| $0.063096 \leq y \leq 0.1$ | 7 | 61 | 6 | $< 1\times10^{-16}$ |

## 4. Results and discussion

4.1 Error analysis with multi-precision computing platform

Table 1 shows that the accuracy of evaluation scheme (27) can reach $1.9055\times10^{-44}$ when $1\times10^{-100} \leq y \leq 0.1$, and the limit accuracy is better than $1\times10^{-100}$ when $1\times10^{-100} \leq y < 1.5849\times10^{-4}$, which indicates that this evaluation scheme can provide highly accurate values of Voigt/complex error function with small imaginary argument against which the accuracy achieved by fast algorithms maybe benchmarked. By contrast, Boyer and Lynas-Gray[10] used multi-precision algorithm to calculate the high precision benchmark value of Voigt/complex error function with a maximum absolute error in both Re $[w(z)]$ and Im $[w(z)]$ of $10^{-100}$ for $|z| \leq 40$. However, Boyer and Lynas-Gray's algorithm requires a high number of calculation bits (2000 bits) to reduce the cancellation and smearing errors, since the first 5000 terms of a series need to be calculated.

4.2 Error analysis with double-precision computing platform

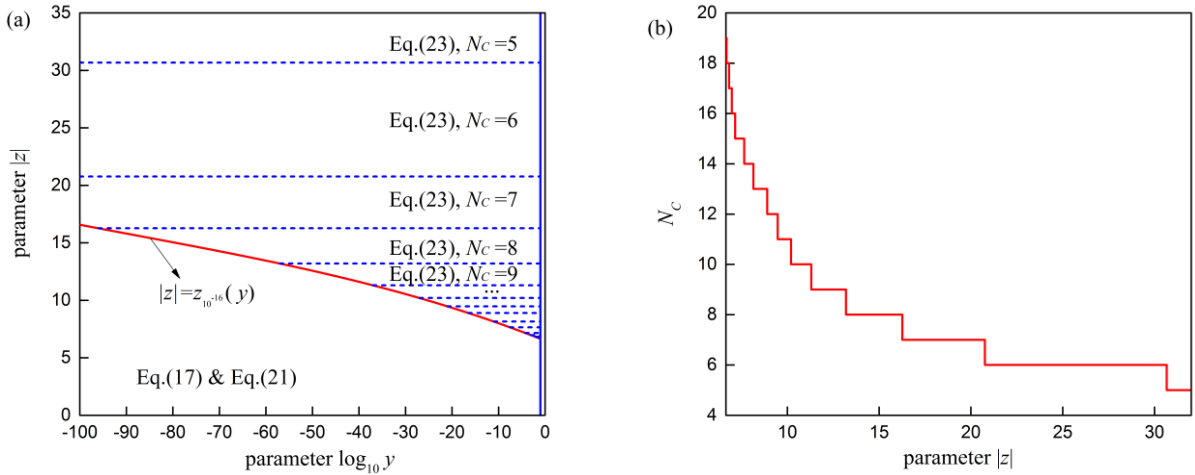

Fig. 3. (a) The evaluation scheme for Voigt/complex error function in double-precision computing platform and (b) the parameter $N_C$ for truncation of the Laplace continued fraction in the external calculation domain.

Since the double-precision floating-point arithmetic commonly used in many science computing and engineering applications, the error analysis and run-time tests of evaluation scheme (26) with double-precision computing platform is needed. In order to quantify accuracy of evaluation scheme (26), it is convenient to define the relative error of Re $[w(z)]$ and Im $[w(z)]$ as

$$\Delta_{Re}(z) = \left| \frac{Re[w(z)] - Re[w_{ref.}(z)]}{Re[w_{ref.}(z)]} \right|,$$
$$\Delta_{Im}(z) = \left| \frac{Im[w(z)] - Im[w_{ref.}(z)]}{Im[w_{ref.}(z)]} \right|. \quad (28)$$

Since the significant figure of double-precision calculation is 16 bits, the efficient computing boundary for $\varepsilon=10^{-16}$ in Table 1 is selected to distinguish the internal and external calculation domain (see Fig. 3a). The parameters $N$ and $N_D$ in the internal calculation domain are shown in Table 3 and the parameter $N_C$ for truncation of the Laplace continued fraction in the external calculation domain is shown in Fig. 3b.

Fig. 4 shows $\log_{10}\Delta_{Re}$ for the real part of the complex error function computed over the domain $0 \leq x \leq 40{,}000 \cap 10^{-100} \leq y \leq 0.1$. As we can see from this figure, the evaluation scheme (26) provides accuracy better than $10^{-15}$ (green color) over the most of this domain. Although accuracy deteriorates in the neighborhood of the efficient computing boundary (see Fig. 4a), it remains better than $10^{-13}$ (red color).

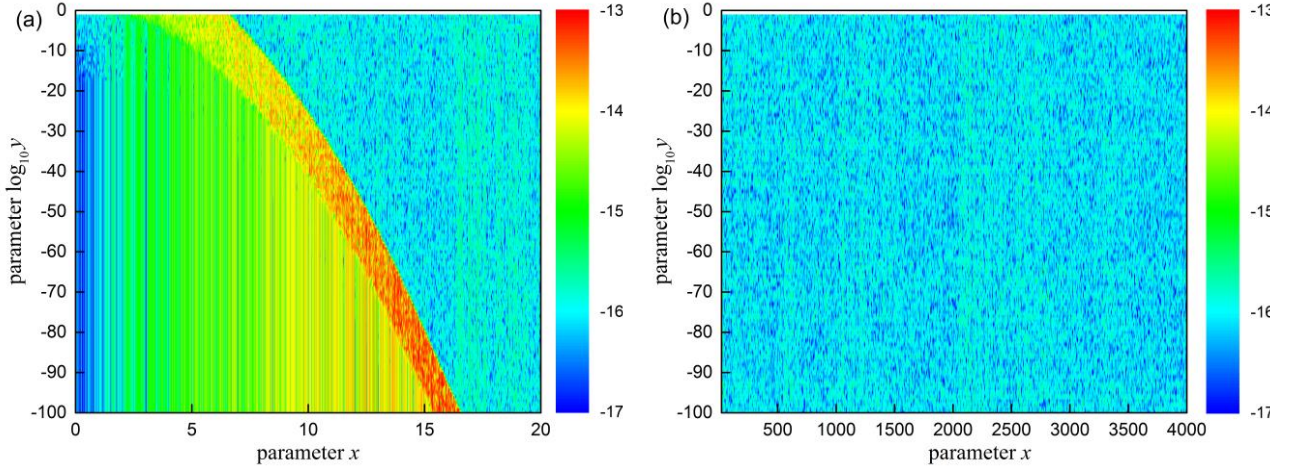

Fig. 4. The logarithm of the relative error $\log_{10}\Delta_{Re}$ for the real part of the evaluation scheme (26) over the domain (a) $0 \leq x \leq 20 \cap 10^{-100} \leq y \leq 0.1$ and (b) $20 \leq x \leq 40{,}000 \cap 10^{-100} \leq y \leq 0.1$, respectively.

Fig. 5 illustrates $\log_{10}\Delta_{Im}$ for the imaginary part of the complex error function also computed over the domain $0 \leq x \leq 40{,}000 \cap 10^{-100} \leq y \leq 0.1$. As we can see from this figure, the evaluation scheme (26) provides accuracy better than $10^{-16}$ (blue color) over the most domain. Unlike the approximation for the real part, the evaluation scheme (26) also highly accurate in the neighborhood of the efficient computing boundary and the computational test reveals that the worst accuracy is better than $10^{-15}$ (green color). One can see that the accuracy of the imaginary part is at least one order of magnitude better than the real part, which can be explained from the fact that the Eq. 21 is derived from the Eq. 17 and derivative operation enlarges calculation error.

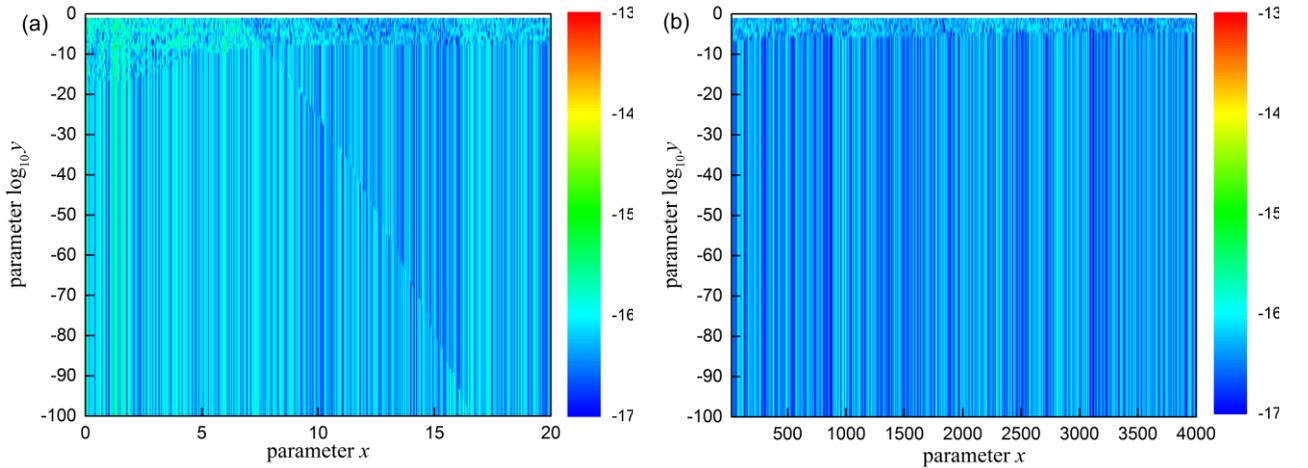

Fig. 5. The logarithm of the relative error $\log_{10}\Delta_{Im}$ for the imaginary part of the evaluation scheme (26) over the domain (a) $0 \leq x \leq 20 \cap 10^{-100} \leq y \leq 0.1$ and (b) $20 \leq x \leq 40{,}000 \cap 10^{-100} \leq y \leq 0.1$, respectively.

In Fig. 6a we compare the maximum relative error $e_{Re}(y) = \max\{\Delta_{Re}(x+iy) \mid 0 \leq x \leq 4000\}$ as a function of $y$ for the real part of complex error function. Obviously, the calculation accuracy of "fexp" algorithm deteriorates further with decreasing $y$ and the maximum relative error even reaches 100% when $y < 10^{-14}$. The calculation accuracy of "voigtf" algorithm is better than $10^{-9}$ when $y > 10^{-5}$, however, similar to "fexp" algorithm, the calculation error increases rapidly with the decrease of $y$ when $y < 10^{-5}$. When $y > 10^{-12}$, "fadsamp" algorithm can achieve $10^{-13}$ calculation accuracy, however, when $y < 10^{-12}$, the accuracy of this algorithm is also fails. It can be seen that these algorithms have their own advantages and disadvantages, has its specific scope, and also needs further improvement. The common point of these algorithms is that they are not suitable for the high precision evaluation of the real part of complex error function with small imaginary argument. Compared with "fexp", "voigtf" and "fadsamp" algorithms, our new algorithm base on evaluation scheme (26) has been significantly improved. The calculation error of the proposed algorithm in this work increases slightly with the decrease of $y$ and the maximum relative error $e_{Re}(y)$ in the domain $10^{-100} \leq y \leq 0.1$ is better than $2.93 \times 10^{-13}$.

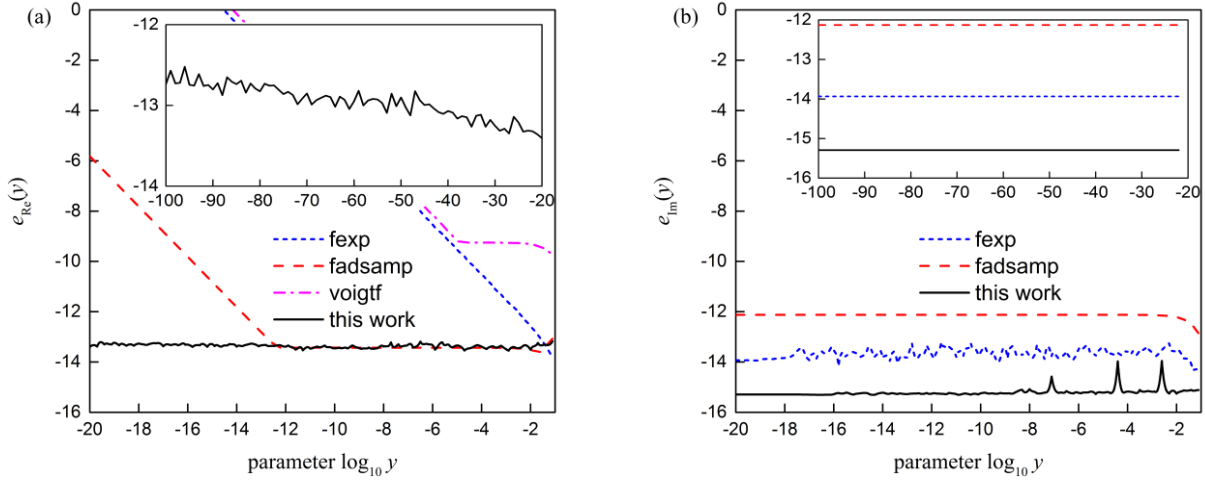

Fig. 6. The maximum relative error for the (a) real part and (b) imaginary part of complex error function.

In Fig. 6b we compare the maximum relative error $e_{Im}(y) = \max\{\Delta_{Im}(x+iy) \mid 0 \leq x \leq 4000\}$ as a function of $y$ for the imaginary part of complex error function. Fortunately, "fexp" algorithm and "fadsamp" algorithm, as well as the proposed algorithm in this work, can achieve high-precision calculation for the imaginary part of complex error function. Compared with the existing algorithms, the proposed algorithm has higher accuracy and the average value of the maximum relative error $e_{Im}(y)$ in the domain $10^{-100} \leq y \leq 0.1$ is $4.91 \times 10^{-16}$.

4.3 Preliminary run-time tests with MATLAB

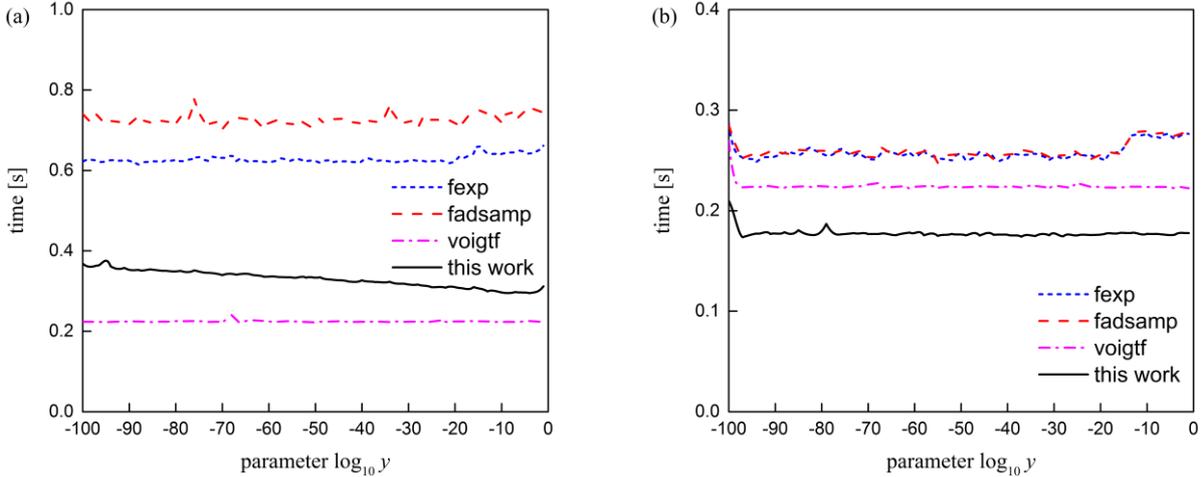

Fig. 7. Execution time (s) per algorithm evaluation with 1 million random numbers in different $y$ within the (a) internal domain $|z| < 22$ and (b) external domain $22 \leq |z| \leq 4000$, respectively. The tests have been performed on a typical desktop computer Intel(R) Quad CPU with RAM 8.00 GB. (Since the "voigtf" algorithm can only calculate the real part of the complex error function, the calculation time is doubled here to compare the calculation speed with other algorithms.)

To get a first impression about the performance of the various approximations, we have used MATLAB's built-in function "tic/toc" to evaluate the calculation speed of these algorithms. Fig. 7 summarizes the results of these numerical experiments. Computational test reveals that with 1 million random numbers in different $y$ within the internal domain $|z| < 22$, the proposed algorithm in this work is faster than "fexp" and "fadsamp" by a factor 1.91 and 2.21, respectively (the average computing time of these algorithms is 0.33 seconds, 0.63 seconds and 0.73 seconds, respectively) and the "voigtf" algorithm (0.22 seconds) is the fastest. In the external domain $22 \leq |z| \leq 4000$, the proposed algorithm has the highest computational efficiency and the calculation speed is faster than "fexp", "fadsamp" and "voigtf" algorithm by a factor 1.44, 1.44 and 1.22, respectively (the average computing time of these algorithms is 0.18 seconds, 0.26 seconds, 0.26 seconds and 0.22 seconds respectively). The preliminary run-time tests show that the proposed algorithm is as fast as that of reported in recent publications.

## 5. Conclusion

In this work we present an efficient approximation for the Voigt/complex error function that can be used for computation at small $y \leq 0.1$. Error analysis and run-time tests with double-precision computing platform reveals that in the real and imaginary parts the proposed algorithm provides average accuracy exceeding $10^{-15}$ and $10^{-16}$, respectively, and the calculation speed is as fast as that of reported in recent publications. Since the evaluation scheme (26) or (27) is a general expression, the desired accuracy can be easily achieved by choosing reasonable parameters. In particular, the optimal parameters under different accuracy levels $\Delta=1\times10^{-100}$ and $\Delta=1\times10^{-16}$ of evaluation scheme (27) are shown in Table 2 and Table 3, respectively, in order to gain computational acceleration. An optimized MATLAB code base on this work can be downloaded from MATLAB Central website: https://www.mathworks.com/matlabcentral/fileexchange/75134-computing-of-voigt-complex-error-function-with-small-y.

## Appendix A

The 0-order and 2-order partial derivatives of $L(x, y)$ with respect to $y$ are expressed in terms of the Dawson's integral $D(x)$ of real argument $x$ as follow:

$$L(x,0) = \frac{2}{\sqrt{\pi}} D(x), \tag{A-1}$$

$$\frac{\partial^2 L(x,0)}{\partial y^2} = \frac{1}{\sqrt{\pi}}((4-8x^2)D(x)+4x). \tag{A-2}$$

We note that the even-order partial derivatives of $L(x, y)$ with respect to $y$ satisfies the following recurrence relation:

$$\frac{\partial^{2m} L(x,0)}{\partial y^{2m}} = (8m-6-4x^2)\frac{\partial^{2m-2}L(x,0)}{\partial y^{2m-1}} - 8(m-1)(2m-3)\frac{\partial^{2m-4}L(x,0)}{\partial y^{2m-4}}, m \geq 2. \tag{A-3}$$

Thus Eq. (9) ~ (14) can be easily derived from the above equation. It should be noted that the coefficients $p_{n/2,k}$ have the following simple expression:

$$p_{n/2,k} = \frac{(-1)^k n! 2^{k+1}}{(n/2-k)!k!(2k-sign(k))!!}. \tag{A-4}$$

However, we did not find a simple expression for the coefficients $q_{n/2,k}$. Instead, the coefficients $q_{n/2,k}$ ($N \leq 7$) is given as shown in the following table A-1 according to the recurrence relation.

Table A-1. The coefficients $q_{n/2,k}$ ($N \leq 7$).

|        | k=0       | k=1        | k=2       | k=3       | k=4      | k=5     | k=6   |
|--------|-----------|------------|-----------|-----------|----------|---------|-------|
| n/2=0  | 0         | 0          | 0         | 0         | 0        | 0       | 0     |
| n/2=1  | 4         | 0          | 0         | 0         | 0        | 0       | 0     |
| n/2=2  | 40        | -16        | 0         | 0         | 0        | 0       | 0     |
| n/2=3  | 528       | -448       | 64        | 0         | 0        | 0       | 0     |
| n/2=4  | 8928      | -11840     | 3456      | -256      | 0        | 0       | 0     |
| n/2=5  | 185280    | -337920    | 150528    | -22528    | 1024     | 0       | 0     |
| n/2=6  | 4567680   | -10671360  | 6429696   | -1456128  | 133120   | -4096   | 0     |
| n/2=7  | 130556160 | -373416960 | 284691456 | -86630400 | 11939840 | -737280 | 16384 |